\documentclass[aps]{revtex4}
\usepackage{amsfonts}
\usepackage{amsmath}
\usepackage{amssymb}
\usepackage{graphicx}
\usepackage{bm}

\input{tcilatex}

\begin{document}

\title{Thermally driven Marangoni surfers}
\author{Alois W\"{u}rger}
\affiliation{Laboratoire Ondes et Mati\`{e}re d'Aquitaine, Universit\'{e} de Bordeaux \&
CNRS, 33405 Talence, France}

\begin{abstract}
We study auto-propulsion of a interface particle, which is driven by the
Marangoni stress arising from a self-generated asymmetric temperature or
concentration field. We calculate separately the long-range Marangoni flow $%
\mathbf{v}^{I}$ due to the stress discontinuity at the interface and the
short-range velocity field\textbf{\ }$\mathbf{v}^{P}$ imposed by the
no-slip condition on the particle surface; both contributions are evaluated
for a spherical floater with temperature monopole and dipole moments. We
find that the self-propulsion velocity is given by the amplitude of the
\textquotedblleft source doublet\textquotedblright\ which belongs to
short-range contribution\textbf{\ }$\mathbf{v}^{P}$. Hydrodynamic
interactions, on the other hand, are determined by the long-range Marangoni
flow $\mathbf{v}^{I}$; its dipolar part results in an asymmetric
advection pattern of neighbor particles, which in turn may perturb the known
hexatic lattice or even favor disordered states.
\end{abstract}

\maketitle

\section{Introduction}

Autonomous motion is an important issue in active soft matter, with possible
applications ranging from energy harvesting \cite{Oka09} to microfluidic
transport \cite{Del12} and mixing \cite{Ven13}. Recently realized
microswimmers carry, as active element, heat-absorbing or catalytic parts
which generate temperature or concentration gradients in the surrounding
liquid.\ For particles dispersed in bulk phases, these thermodynamic forces
give rise to an effective slip velocity, which in turn implies
self-propulsion at a speed of the order of ten microns per second \cite%
{Pax05,Qia13}.

Much higher velocities can be achieved for active particles trapped at a
liquid interface, where the self-generated temperature or concentration
gradient induces a non-uniform interface tension and a Marangoni flow. Upon
laser-heating one side of a centimeter-size object floating on water, \cite%
{Oka09} observed self-propulsion at several cm/s. The mutual Marangoni
advection of camphor releasing gel particles results in dynamical self
assembly \cite{Soh08}

\FRAME{ftbpFU}{3.8562in}{4.6743in}{0pt}{\Qcb{Schematic view of a trapped
particle with an off-center active spot. a) The Marangoni flow is symmetric
with respect to the hot spot, yet not with respect to the particle center.
b) Side view of a particle at a fluid interface separating phases 1 and 2
with thermal conductivity $\protect\kappa _{i}$ and viscosity $\protect\eta %
_{i}$. The polar angle $\protect\theta $ is defined with respect to the $z$%
-axis. c) Top view with the azimuthal angle $\protect\varphi $. The
idealized heat source is indicated by the black point at a distance $b$ from
the particle center. \ }}{}{fig.tiff}{\special{language "Scientific
Word";type "GRAPHIC";maintain-aspect-ratio TRUE;display "USEDEF";valid_file
"F";width 3.8562in;height 4.6743in;depth 0pt;original-width
6.7914in;original-height 8.2399in;cropleft "0";croptop "1";cropright
"1";cropbottom "0";filename 'Fig.TIFF';file-properties "XNPEU";}}

Recent theoretical studies dealt with Marangoni propulsion due to a
non-uniform surfactant concentration. The resulting Marangoni stress
comprises a dipolar term that is proportional to the inverse distance from
its source. Thus \cite{Lau12} calculated the Marangoni flow for an
asymmetric active disk, whereas \cite{Mas14} considered the motion of
ellipsoidal particles in terms of a reciprocal theorem. Accounting for
advective non-linearities in the surfactant concentration, \cite{Nag13}
found spontaneous rotation of particles floating on a liquid droplet.

In the present paper we consider Marangoni propulsion driven by\ a
non-uniform temperature or bulk concentration field, the dipolar term of
which decays with the square of inverse distance.\ Assuming small Reynolds,
Marangoni, P\'{e}clet, and capillary numbers, we rely on Stokes' equation,
linearize the field dependent tension, and neglect the interface
deformation. For a spherical particle with an off-center heat source, we
calculate both the long-range Marangoni flow $\mathbf{v}^{I}$ driven by
the non-uniform interface tension and the short-range velocity $\mathbf{v%
}^{P}$ imposed by no-slip condition on the particle surface.

\section{Marangoni effect and boundary conditions}

Consider a colloidal particle trapped at a liquid interface. For the sake of
simplicity we discuss a sphere that is trapped at midplane, as shown in
Fig.\ 1.\ Retaining the first two terms of the multipole expansion of the
temperature profile we find 
\begin{equation}
T(\mathbf{\emph{r}})=T_{0}+\frac{Q}{2\pi \bar{\kappa}}\left( \frac{1}{r}+%
\frac{\mathbf{b\cdot r}}{r^{3}}+...\right)  \label{temp}
\end{equation}%
where $Q$ is the total power absorbed by the particle and $\bar{\kappa}%
=\kappa _{1}+\kappa _{2}$ the sum of the thermal conductivities of the lower
and upper fluid phases. The strength and orientation of the dipole term are
given by the position of the absorption area with respect to the particle
center, $\mathbf{b}=b\mathbf{e}_{x}$, as shown in Fig.\ 1 for a
point-like heat source. Moreover, $b$ depends on the conductivity contrast
of the particle and fluid phases. In the case of a chemical Marangoni
effect, $T$ and $\kappa $ are the solute concentration and diffusivity in
the fluid phases and $Q$ is the chemical activity at the particle surface.

The thermal gradient induces a Marangoni flow along the liquid interface.
The temperature dependence of the interface tension, $\gamma _{T}=d\gamma
/dT $, results in the hydrodynamic boundary condition 
\begin{equation}
(1-\mathbf{nn)}\cdot \left( \mathbf{\sigma }^{M}\cdot \mathbf{n}%
+\gamma _{T}\mathbf{\nabla }T\right) =0,  \label{2}
\end{equation}%
where $\mathbf{\sigma }^{M}=\mathbf{\sigma }^{(1)}-\mathbf{%
\sigma }^{(2)}$\ is the stress discontinuity across the interface, $%
\mathbf{n}$ the (downward oriented) normal vector, and $\mathbf{%
\nabla }\gamma =\gamma _{T}\mathbf{\nabla }T$ the tension gradient. The
stress tensor $\mathbf{\sigma }$ comprises viscous and pressure terms, $%
\sigma _{ij}=\eta (\partial _{i}v_{j}+\partial _{j}v_{i})-P\delta _{ij}$,
with the velocity components $v_{i}$ and where the viscosity $\eta $ takes
values $\eta _{1}$ and $\eta _{2}$ in the two phases. The fluid velocity is
continuous at the interface, and its normal component must vanish, 
\begin{equation}
\mathbf{v\cdot \mathbf{n}|}_{I}=0.
\end{equation}%
One more boundary condition is provided by the no-slip condition at the
particle surface,%
\begin{equation}
\mathbf{v|}_{P}=\mathbf{u},
\end{equation}%
where $\mathbf{v(r)}$ is the velocity field of the fluid and $%
\mathbf{u}$ the particle velocity, which is necessarily parallel to the
interface.

Since there are no external body forces acting on the particle we have 
\begin{equation}
\oint \mathrm{d}\mathbf{s}\gamma +\oint \mathrm{d}\mathbf{S}\cdot 
\mathbf{\sigma }=0,  \label{6}
\end{equation}%
where the first term gives the net force exerted by the surface tension
along the contact line with element $d\mathbf{s}$, and the second one,
the surface integral of the stress tensor with the oriented surface element $%
d\mathbf{S}$.

For later convenience we rewrite the above conditions for the velocity and
stress compenents in spherical coordinates. With the axes defined in Fig. 1
we have $\mathbf{n}=\mathbf{e}_{\theta }$ and 
\begin{subequations}
\label{12}
\begin{equation}
(1-\mathbf{nn})\cdot \mathbf{\sigma }^{M}\cdot \mathbf{n}=%
\mathbf{e}_{r}\sigma _{r\theta }^{M}+\mathbf{e}_{\varphi }\sigma
_{\varphi \theta }^{M}.
\end{equation}%
Inserting the temperature gradient in (\ref{2}), we find 
\begin{equation}
\sigma _{r\theta }^{M}=\frac{\gamma _{T}Q}{2\pi \bar{\kappa}}\left( \frac{1}{%
r^{2}}+2b\frac{\cos \varphi }{r^{3}}\right) ,\ \ \ \ \sigma _{\varphi \theta
}^{M}=\frac{\gamma _{T}Q}{2\pi \bar{\kappa}}\frac{b\sin \varphi }{r^{3}}.
\end{equation}%
The zero normal velocity at the interface involves the polar component only, 
\end{subequations}
\begin{equation}
v_{\theta }|_{I}=0.  \label{13}
\end{equation}%
Finally, imposing no-slip at the surface of the particle moving at velocity $%
u$ along the $x$-axis, we have 
\begin{equation}
v_{r}|_{P}=u\sin \theta \cos \varphi ,\ \ \ v_{\theta }|_{P}=u\cos \theta
\cos \varphi ,\ \ \ v_{\varphi }|_{P}=-u\sin \varphi .  \label{14}
\end{equation}

The cosine $\cos \theta $ turns out to be a more convenient coordinate than
the polar angle $\theta $. Thus we use the shorthand notation 
\begin{equation*}
c=\cos \theta ,\ \ \ s=\sqrt{1-c^{2}}.
\end{equation*}%
Note that $s$ is positive everywhere, whereas $c>0$ and $c<0$ on the upper
and lower halfspaces, respectively.

Finally we note that the particle is not allowed to rotate about the
horizontal axis. In other words, the point $b$ in Fig.\ 1 is confined to the
interface plane. This condition could be imposed through appropriate surface
functionalization; it is satisfied, for example, if heating occurs through a
small metal patch which itself is trapped at the fluid phase boundary.

\section{Velocity field and stress tensor}

The velocity field is solution of Stokes' equation $\eta \mathbf{\nabla }%
^{2}\mathbf{v}=\mathbf{\nabla }P$, which results in a set of coupled
differiential equations for the three velocity components $\mathbf{v}%
=v_{r}\mathbf{e}_{r}+v_{\theta }\mathbf{e}_{\theta }+v_{\varphi }%
\mathbf{e}_{\varphi }$ and the pressure $P$. With the vector Laplace
operator one has 
\begin{subequations}
\label{1}
\begin{eqnarray}
\Delta v_{r}-\frac{2v_{r}}{r^{2}}+\frac{2\partial _{c}(sv_{\theta })}{r^{2}}-%
\frac{2\partial _{\varphi }v_{\varphi }}{sr^{2}} &=&\frac{\partial _{r}P}{%
\eta }, \\
\Delta v_{\theta }-\frac{v_{\theta }}{s^{2}r^{2}}-\frac{2s\partial _{c}v_{r}%
}{r^{2}}-\frac{2c\partial _{\varphi }v_{\varphi }}{s^{2}r^{2}} &=&-s\frac{%
\partial _{c}P}{\eta r},\ \ \ \  \\
\Delta v_{\varphi }-\frac{v_{\varphi }}{s^{2}r^{2}}+\frac{2\partial
_{\varphi }v_{r}}{sr^{2}}+\frac{2c\partial _{\varphi }v_{\theta }}{s^{2}r^{2}%
} &=&\frac{\partial _{\varphi }P}{\eta sr},
\end{eqnarray}%
where the derivative with respect to the polar angle is replaced according
to $\partial _{\theta }=-s\partial _{c}$. The first term of each equation is
given by the scalar Laplace operator 
\begin{equation}
\Delta v_{i}=r^{-2}\partial _{r}(r^{2}\partial _{r}v_{i})+r^{-2}\partial
_{c}(s^{2}\partial _{c}v_{i})+(sr)^{-2}\partial _{\varphi }^{2}v_{i}.
\end{equation}%
Similarly, we have for the incompressibility condition 
\begin{equation}
\mathbf{\nabla \cdot v}=r^{-2}\partial _{r}(r^{2}v_{r})-r^{-1}\partial
_{c}(sv_{\theta })+(sr)^{-1}\partial _{\varphi }v_{\varphi }=0.
\end{equation}%
For later use we give the components of the symmetrized stress in spherical
coordinates, 
\end{subequations}
\begin{align}
\sigma _{rr}& =2\eta \partial _{r}v_{r}-P,\ \ \ \ \sigma _{\theta \theta
}=(2\eta /r)\left( -\partial _{c}v_{\theta }+v_{r}\right) -P,  \notag \\
\sigma _{\varphi \varphi }& =(2\eta /sr)\left( \partial _{\varphi
}v_{\varphi }+sv_{r}+cv_{\theta }\right) -P,\ \ \sigma _{r\theta }=(\eta
/r)\left( r\partial _{r}v_{\theta }-v_{\theta }-s\partial _{c}v_{r}\right) ,
\label{3} \\
\sigma _{r\varphi }& =(\eta /sr)\left( sr\partial _{r}v_{\varphi
}-sv_{\varphi }+\partial _{\varphi }v_{r}\right) ,\ \ \sigma _{\varphi
\theta }=(\eta /sr)\left( -s^{2}\partial _{c}v_{\varphi }+\partial _{\varphi
}v_{\theta }-cv_{\varphi }\right) .  \notag
\end{align}%
In view of (\ref{2}) we need the off-diagonal components $\sigma _{r\theta }$
and $\sigma _{\varphi \theta }$. The above relations are valid in each of
the fluid phases, albeit with different viscosities $\eta _{1}$ and $\eta
_{2}$.

We write the fluid velocity as the sum of two terms, 
\begin{equation}
\mathbf{v}=\mathbf{v}^{I}+\mathbf{v}^{P},
\end{equation}%
the first of which accounts for the source field (\ref{12}) with the
boundary condition (\ref{13}). This field fully describes the Marangoni flow
induced by a pointlike heat source. For a particle of finite size, the
second term $\mathbf{v}^{P}$ is required in order to assure the no-slip
condition at its surface.\ This is achieved by chosing $\mathbf{v}^{P}$
such that it satisfies both (\ref{13}) and (\ref{14}), yet does not
contribute to the Marangoni stress.

\section{Interface contribution $v^{I}$}

Since the Marangoni stress (\ref{12}) results in a jump of the velocity
derivatives, the velocity is not analytic at the interface. The isotropic
component of $\sigma ^{M}$ varies with the distance as $1/r^{2}$, and the
dipolar contribution as $1/r^{3}$. This implies that $\mathbf{v}^{I}$
consists of axially symmetric and dipolar terms proportional to $1/r$ and $%
1/r^{2}$, respecively.

\subsection{Axisymmetric part}

The Marangoni flow of a heat source at the origin is obtained from the
stream function $\psi =-Uacr(1\mp c)$ with the velocity scale $U$, 
\begin{equation}
v_{r}^{I}=\frac{\partial _{\theta }\psi }{sr^{2}}=U\frac{a}{r}\left( 1\mp
2c\right) ,\ \ \ v_{\theta }^{I}=-\frac{\partial _{r}\psi }{sr}=U\frac{a}{r}%
\frac{cs}{1\pm c}.  \label{22}
\end{equation}%
Both radial and polar components are proportional to the inverse distance; $%
v_{\theta }^{I}$ diverges on the positive $z$-axis for the minus sign, and
on the negative one for the plus sign.

\subsection{Dipolar part}

We look for solutions of Stokes' equation that vary with distance as $r^{-2}$
and that are singular at $c=\pm 1$. With the coordinates defined in Fig.\ 1,
it is clear that that radial and polar components are proportional to $\cos
\varphi $, and $v_{\varphi }^{I}\propto \sin \varphi $. Thus we have 
\begin{equation}
v_{r}^{I}=f_{r}(c\mathcal{)}\frac{\cos \varphi }{r^{2}},\ \ v_{\theta
}^{I}=f_{\theta }(c)\frac{\cos \varphi }{r^{2}},\ \ v_{\varphi
}^{I}=f_{\varphi }(c\mathcal{)}\frac{\sin \varphi }{r^{2}},\ \ P^{I}=f_{P}(c%
\mathcal{)}\frac{\cos \varphi }{r^{3}}.
\end{equation}%
Inserting this ansatz in (\ref{1}) results in four coupled equations for the
functions $f_{i}(c)$; the incompressibility condition involves $f_{\theta }$
and $f_{\varphi }$\ only.\ 

By successive elimination we obtain a single fourth-order differential
equation for $f_{\theta }$, 
\begin{equation}
\partial _{c}^{3}(1-c^{2})^{2}\partial _{c}f_{\theta }=0.
\end{equation}%
From its solution one readily constructs the remaining components and the
related pressure; discarding regular and logarithmic terms we have 
\begin{equation}
f_{\theta }=\frac{t_{2}+t_{3}c}{1-c^{2}},\ \ \ f_{r}=\frac{t_{2}c^{3}+t_{3}}{%
\sqrt{1-c^{2}}},\ \ \ f_{\varphi }=\frac{t_{2}c+t_{3}}{1-c^{2}},\ \ \ \
f_{P}=2t_{2}c\sqrt{1-c^{2}}  \label{34}
\end{equation}%
with coefficients $t_{2}$ and $t_{3}$. Note that the velocity field diverges
along the vertical axis $c=\pm 1$.\ 

\subsection{Velocity field and stress in the upper and lower halfspaces}

In view of the stress jump (\ref{12}) it is clear that the velocity above
and below the interface cannot be given in closed form. Analytic solutions
in the halfspaces $c\gtrless 0$ are obtained by putting 
\begin{equation*}
t_{2}=\mp t_{3}.
\end{equation*}%
Then the singular factors in (\ref{34}) disappear, and the relevant solution
is given by 
\begin{subequations}
\begin{eqnarray}
v_{r}^{I} &=&U\frac{1\mp 2c}{\hat{r}}+\left( t_{3}\frac{1\mp c^{3}}{s}\mp
t_{4}sc\right) \frac{\cos \varphi }{\hat{r}^{2}}, \\
v_{\theta }^{I} &=&U\frac{sc}{\hat{r}(1\pm c)}\mp \left( \frac{t_{3}}{1\pm c}%
-t_{5}\right) \frac{\cos \varphi }{\hat{r}^{2}}, \\
v_{\varphi }^{I} &=&\left( \frac{t_{3}}{1\pm c}\mp t_{5}c\right) \frac{\sin
\varphi }{\hat{r}^{2}},
\end{eqnarray}%
with $\hat{r}=r/a$ and where the plus and minus signs are valid in the upper
and lower halfspaces, respectively. The axisymmetric part corresponds to (%
\ref{22}), whereas the terms proportional to $t_{3}$ are obtained from (\ref%
{34}). The contributions with $t_{4}$\ and $t_{5}$\ are well-known regular
flow patterns \cite{Sch82} that have been added in order to meet the
boundary conditions at the interface.

The velocity field is symmetric with respect to the interface plane and
continuous at $c=0$, yet its derivatives are not and lead to a stress
discontinuity along the interface.Finally, the corresponding pressure reads
as 
\begin{equation}
P^{I}=\mp 2\eta a\left( U\frac{c}{r^{2}}+(t_{4}-t_{3})a\frac{sc}{r^{3}}\cos
\varphi \right) .
\end{equation}%
The pressure vanishes at $c=0$ and thus is continuous at the interface.
Since the viscosity $\eta $ in general takes different values in the upper
and lower phases, the normal component of the pressure gradient is
discontinuous.

Now we determine the parameters $U$ and $t_{i}$ from the boundary
conditions. In view of (\ref{13}) for the normal velocity component we put $%
c=0$ and find 
\end{subequations}
\begin{equation*}
t_{5}=t_{3}.
\end{equation*}%
From (\ref{3}) we calculate the off-diagonal stress components $\sigma
_{r\theta }$ and $\sigma _{\varphi \theta }$\ at both sides of the
interface. Their difference $\mathbf{\sigma }^{M}=\mathbf{\sigma }%
^{(1)}-\mathbf{\sigma }^{(2)}$ is proportional to the sum of the
viscosities $\bar{\eta}=\eta _{1}+\eta _{2}$, 
\begin{equation*}
\sigma _{r\theta }^{M}=-2\bar{\eta}a\frac{U}{r^{2}}-\bar{\eta}a^{2}\frac{%
t_{4}}{r^{3}},\ \ \ \ \ \sigma _{\theta \varphi }^{M}=-2\bar{\eta}a^{2}\frac{%
t_{3}}{r^{3}}.
\end{equation*}%
Inserting the temperature gradient in the boundary conditions (\ref{12}),
one readily obtains the velocity scale $U$ and the remaining coefficients $%
t_{i}$, 
\begin{equation}
U=-\frac{\gamma _{T}Q}{4\pi \bar{\eta}\bar{\kappa}a},\ \ \ t_{3}=\frac{b}{a}%
U,\ \ \ t_{4}=4\frac{b}{a}U.  \label{15}
\end{equation}%
(For a liquid-air interface one has $\bar{\eta}=\eta _{1}$ and $\bar{\kappa}%
=\kappa _{1}$.) Then the velocity field reads 
\begin{subequations}
\label{5}
\begin{eqnarray}
v_{r}^{I} &=&U\frac{a}{r}\left( 1\mp 2c+\frac{b}{r}\frac{1\mp c^{3}\mp
4cs^{2}}{s}\cos \varphi \right) ,\ \ \ \  \\
v_{\theta }^{I} &=&U\frac{a}{r}\left( \frac{cs}{1\pm c}+\frac{b}{r}\frac{c}{%
1\pm c}\cos \varphi \right) , \\
v_{\varphi }^{I} &=&U\frac{ab}{r^{2}}\left( \frac{1}{1\pm c}\mp c\right)
\sin \varphi ,
\end{eqnarray}%
where the upper and lower signs occur in the upper and lower halfspaces,
respectively. The corresponding pressure 
\end{subequations}
\begin{equation}
P^{I}=\mp 2aU\eta \left( \frac{c}{r^{2}}+3b\frac{sc}{r^{3}}\cos \varphi
\right)  \label{5a}
\end{equation}%
is proportional to $c=\cos \theta $ and thus vanishes at the interface.

Eqs. (\ref{5}) and (\ref{5a}) constitute the solution for a pointlike
interface particle carrying temperature monopole and dipole moments. The
latter results in a radial flow velocity at the interface proportional to $%
r^{-2}\cos \varphi $.

\subsection{Stress and force balance}

From elementary symmetry considerations it is clear that both contributions
to (\ref{6}) have finite components along the $x$-axis only. Thus the
contact line element $\mathrm{d}\mathbf{s}$ reduces to the component $%
\mathrm{d}s_{x}=a\mathrm{d}\varphi \cos \varphi $, and the non-uniform
interface tension leads to a net force $\oint \mathrm{d}s_{x}\gamma $. The
drag force exerted on the surface element $dS=a^{2}dcd\varphi $ is given by
the stress component $\sigma _{rx}^{I}$; from (\ref{3}) and (\ref{5}) one
finds 
\begin{equation*}
\frac{\sigma _{rx}^{I}}{\eta U/a}=\left( \frac{-2s}{1\pm c}\pm 6sc\right)
\cos \varphi +\frac{b}{a}\left( \frac{-4}{1\pm c}\pm 22c\mp 24c^{3}\right)
\cos ^{2}\varphi +\frac{b}{a}\left( \frac{4}{1\pm c}\mp 6c\right) \sin
^{2}\varphi .
\end{equation*}%
Integrating the linearized tension $\gamma =\gamma _{0}+\gamma _{T}(T-T_{0})$
along the contact line and the stress over the particle surface, we find
that the tension and drag forces cancel each other, 
\begin{equation}
\oint \mathrm{d}s_{x}\gamma +\oint \mathrm{d}S\sigma _{rx}^{I}=-2\pi \bar{%
\eta}bU+2\pi \bar{\eta}bU=0.  \label{18}
\end{equation}

\section{Particle contribution $\mathbf{v}^{P}$}

Now we turn to the additional velocity $\mathbf{v}^{P}$ which is
required in order to satisfy the stick boundary condition (\ref{14}) at the
particle surface, 
\begin{equation}
\mathbf{u}=\mathbf{v}^{I}\mathbf{|}_{P}+\mathbf{v}^{P}%
\mathbf{|}_{P}.  \label{19}
\end{equation}%
In addition, $\mathbf{v}^{P}$ has continuous derivatives\ and its normal
component vanishes at the phase boundary.\ In other words, $\mathbf{v}%
^{P}$ does not contribute to the Marangoni stress (\ref{12}) and satisfies
the condition (\ref{13}).

We start from the general analytic solution of Stokes' equation (\ref{1})
which is given by the well-known series \cite{Sch82} 
\begin{subequations}
\label{20}
\begin{align}
v_{r}^{P}& =U\sum_{n,m}\frac{a^{n}}{r^{n}}\left( p_{nm}+q_{nm}\frac{a^{2}}{%
r^{2}}\right) X_{nm}, \\
v_{\theta }^{P}& =U\sum_{n,m}\left[ \frac{a^{n}}{r^{n}}\left( \frac{n-2}{%
n(n+1)}p_{nm}+\frac{q_{nm}}{n+1}\frac{a^{2}}{r^{2}}\right) s\partial
_{c}X_{nm}+\frac{a^{n+1}}{r^{n+1}}\,s_{nm}\frac{\partial _{\varphi }X_{nm}}{s%
}\right] \\
v_{\varphi }^{P}& =U\sum_{n,m}\left[ -\frac{a^{n}}{r^{n}}\left( \frac{n-2}{%
n(n+1)}p_{nm}+\frac{q_{nm}}{n+1}\frac{a^{2}}{r^{2}}\right) \frac{\partial
_{\varphi }X_{nm}}{s}+\frac{a^{n+1}}{r^{n+1}}\,s_{nm}s\partial _{c}X_{nm}%
\right]
\end{align}%
where $n=1,2,3,...$ and $-n\leq m\leq n$. The angle-dependent basis
functions 
\end{subequations}
\begin{equation*}
X_{nm}(c,\varphi )=\left\{ 
\begin{array}{ccc}
P_{n}^{m}(c)\cos (m\varphi ) & \text{for} & m\geq 0, \\ 
P_{n}^{m}(c)\sin (m\varphi ) & \text{for} & m<0,%
\end{array}%
\right.
\end{equation*}%
involve associated Legendre polynomials,%
\begin{equation*}
P_{n}^{m}(c)=\frac{(-s)^{m}}{2^{n}n!}\partial _{c}^{n+m}(c^{2}-1)^{n},
\end{equation*}%
for example $P_{1}^{1}=-s$, $P_{1}^{-1}=s/2$, and $P_{2}^{1}=3sc$. The
inhomogeneous solutions with coefficients $p_{nm}$\ are related to the
pressure\ 
\begin{equation}
P^{P}=2\eta U\sum_{n,m}\frac{a^{n}}{r^{n+1}}\frac{2n-1}{n+1}p_{nm}X_{nm},
\end{equation}%
Since the source comprises monopole and dipole moments only, the above
series has finite terms with $n=1,2,3,...$ and $m=-1,0,1$.

\subsection{Drag force and interface boundary conditions}

The only components that contribute to the drag force are the Stokeslet
terms $p_{1m}$, with $m=-1,0,1$. Here we are interested in the force along
the $x$-axis and thus retain the coefficient $p_{11}$ only. \ The relevant
component of the stress tensor reads 
\begin{equation*}
\sigma _{rx}^{P}=3p_{11}(\eta U/a)\left( 1-c^{2}\right) \cos ^{2}\varphi .
\end{equation*}%
In view of (\ref{6}) and (\ref{18}), it is clear that the surface integral
of $\sigma _{rx}^{P}$ vanishes, 
\begin{equation}
\oint \mathrm{d}S\sigma _{rx}^{P}=2\pi p_{11}\bar{\eta}aU=0.  \label{21}
\end{equation}%
In physical terms this means that there is no additional drag force arising
from $\sigma _{rx}^{P}$ and thus no Stokeslet contribution to the flow $%
\mathbf{v}^{P}$.

In order not to interfere with the above solution $\mathbf{v}^{I}$ and
the conditions (\ref{12}) and (\ref{5}), the additional term $\mathbf{v}%
^{P}$ satisfies 
\begin{equation}
v_{\theta }^{P}|_{I}=0,\ \ \sigma _{r\theta }^{P}|_{I}=0,\ \ \ \sigma
_{\theta \varphi }^{P}|_{I}=0.
\end{equation}%
Putting $c=0$ in both velocity and stress components, one finds that all
coefficients vanish except for \ 
\begin{equation*}
p_{2n,0},\ \ q_{2n,0},\ \ p_{2n+1,1},\ \ q_{2n-1,1},\ \ s_{2n,-1},
\end{equation*}%
with $n=1,2,3,...$ These coefficients are determined from the boundary
condition at the particle surface (\ref{5}).

\subsection{Axisymmetric part}

We consider the velocity contributions that do not depend on the azimuthal
angle $\varphi $. Inserting the velocity field $\mathbf{v}^{P}$ in (\ref%
{19}), we find that the even coefficients $p_{2n,0}$ and $q_{2n,0}$ are
determined by the axisymmetric part of $v_{r}$ and\ $v_{\theta }$, according
to%
\begin{equation}
0=(1\mp 2c)+\sum_{k\text{ even}}\left( p_{k0}+q_{k0}\right) P_{k},\ \ 0=%
\frac{c}{1\mp c}+\sum_{k\text{ even}}\frac{(k-2)p_{k0}+kq_{k0}}{k(k+1)}%
s\partial _{c}P_{k}.
\end{equation}%
Expanding these equations in terms of Legendre polynomials one obtains 
\begin{equation}
q_{2n,0}=\frac{(-1)^{n}(4n+1)(2n)!}{2^{2n+1}n!(n+1)!},\ \ \ \ \ p_{2n,0}=-%
\frac{2n+1}{2n-1}q_{2n,0}.
\end{equation}%
The coefficient $p_{20}$ provides the leading term in inverse powers of
distance, $\mathbf{v}^{P}\propto r^{-2}$. Thus $\mathbf{v}^{P}$
decays faster than the Marangoni flow $\mathbf{v}^{I}\propto r^{-1}$.

\subsection{Dipolar part.}

Now we turn to the dipolar terms $m=\pm 1$ in $\mathbf{v}^{P}$.
Performing the derivatives with respect to $\varphi $, separating factors of 
$\cos \varphi $ and $\sin \varphi $, and dividing by the velocity scale $U$,
we obtain from Eq. (\ref{19}) the relations \ 
\begin{subequations}
\label{23}
\begin{eqnarray}
\frac{u}{U}s &=&\frac{b}{a}\frac{1\mp c^{3}\mp 4cs^{2}}{s}+\sum_{k\text{ odd}%
}\left( p_{k1}+q_{k1}\right) P_{k}^{1}, \\
\frac{u}{U}c &=&\frac{b}{a}\frac{c}{1\pm c}+\sum_{k\text{ odd}}\frac{%
(k-2)p_{k1}+kq_{k1}}{k(k+1)}s\partial _{c}P_{k}^{1}+\sum_{k\text{ even}%
}s_{k,-1}\frac{P_{k}^{-1}}{s}, \\
-\frac{u}{U} &=&\frac{b}{a}\frac{1\mp c-c^{2}}{1\pm c}+\sum_{k\text{ odd}}%
\frac{(k-2)p_{k1}+kq_{k1}}{k(k+1)}\frac{P_{k}^{1}}{s}+\sum_{k\text{ even}%
}s_{k,-1}s\partial _{c}P_{k}^{-1},
\end{eqnarray}%
where the upper and lower signs are valid in the halfspaces $c>0$ and $c<0$,
respectively.

The first equation is solved by expanding\ in terms of the polynomials $%
P_{k}^{1}$, 
\end{subequations}
\begin{equation}
-\frac{u}{U}\delta _{k1}=\frac{b}{a}B_{k}+\left( p_{k1}+q_{k1}\right) \ \ \
(k\text{ odd}).  \label{24}
\end{equation}%
We have used $P_{1}^{1}=-s$ and the coefficients 
\begin{equation*}
B_{2n-1}=\frac{\int \mathrm{d}cP_{2n-1}^{1}(c)[1\mp c^{3}\mp 4cs^{2}]/s}{%
\int \mathrm{d}cP_{2n-1}^{1}(c)^{2}}=\frac{(-1)^{n}(2n)!(8n^{2}-4n-3)}{%
2^{2n}n!(n+1)!(2n-1)(2n-3)},
\end{equation*}%
with $n=1,2,3,...$ The first terms read $B_{1}=\frac{3}{8},$ $B_{3}=\frac{49%
}{96},B_{5}=-\frac{209}{1920}$.

The second and third equations in (\ref{23}) can be conveniently separated
in two parts, one that relates $u/U$ to the coefficients $p_{k1}$ and $%
q_{k1} $, and a second one relating the terms proportional to $b/a$ to the
coefficients $s_{k,-1}$,%
\begin{equation}
\frac{u}{U}\delta _{k1}=\frac{(k-2)p_{k1}+kq_{k1}}{k(k+1)},\ \ \ \ \ \ \ \ \
\ \ 0=-\frac{b}{a}\frac{c}{1\pm c}+\sum_{k\text{ even}}s_{k,-1}\frac{%
P_{k}^{-1}}{s}.  \label{25}
\end{equation}%
For $k\geq 2$ equations (\ref{24}) and (\ref{25}) are readily solved,\ 
\begin{equation}
p_{k1}=-\frac{b}{a}\frac{k}{2}B_{k},\ \ \ q_{k1}=\frac{b}{a}\frac{k-2}{2}%
B_{k}\text{.}
\end{equation}%
The coefficients $s_{2n,-1}$ are determined by multiplying the second
equation in\ (\ref{25}) with $s$ and expanding in terms of $P_{k,-1}$; thus
we obtain 
\begin{equation}
s_{2n,-1}=\frac{b}{a}\frac{(-1)^{n}(2n-2)!(4n+1)}{2^{2n-1}(n-1)!(n+1)!}\text{%
.}
\end{equation}

Now we turn to the case $k=1$. Noting $p_{11}=0$ and $B_{1}=\frac{3}{8}$ one
readily determines the coefficient 
\begin{equation}
q_{11}=-\frac{2}{3}B_{1}\frac{a}{b}=-\frac{a}{4b}
\end{equation}%
and the particle velocity 
\begin{equation}
u=\frac{q_{11}}{2}U=-\frac{b}{8a}U.  \label{29}
\end{equation}%
It is worth noting that the coefficient $q_{11}$ corresponds to the
amplitude of the \textquotedblleft source-doublet\textquotedblright\ of a
diagrammatic expansion \cite{Bla74}. The factor $-\frac{b}{8a}$ could be
obtained equally well from the reciprocal theorem developed by \cite{Mas14}.

\section{Discussion}

\subsection{Velocity field in the surrounding fluid}

The total velocity field consists of two contributions, $\mathbf{v}^{I}$%
\textbf{\ }and\textbf{\ }$\mathbf{v}^{P}$, which describe the long-range
behavior and the flow in the vicinity of the trapped particle, respectively.
The first one is entirely determined by the temperature field (\ref{temp}),
whereas the second one depends on the particle size and shape.\ The dipolar
part of the Marangoni flow (\ref{5}) behaves as $\mathbf{v}^{I}\propto
r^{-2}$, whereas to leading order\textbf{\ }$\mathbf{v}^{P}$ decays as $%
r^{-3}$ and is given by the coefficients $q_{11},p_{31},s_{2,-1}$, which
comprise the \textquotedblleft source-doublet\textquotedblright\ of a
diagrammatic expansion \cite{Bla74}. In an alternative approach, the first
contributions to\textbf{\ }$\mathbf{v}^{P}$ could be constructed in
terms of an image system by successive reflections at the particle surface
and at the interface \cite{Mor10}.

The dipolar component of the Marangoni flow varies with the square of the
inverse distance, $\mathbf{v}^{I}\propto r^{-2}\cos \varphi $. This
power law is related to the fact heat or a molecular solute diffuse in a
bulk phase, and that temperature or concentration obey the 3D diffusion
equation. On the other hand, a surfactant diffuses along the 2D interface
only, and the resulting Marangoni flow behaves as $r^{-1}\cos \varphi $ \cite%
{Lau12}.

Throughout this paper we have assumed that the interface plane coincides
with the particle's midplane, requiring the contact angle $\theta _{0}=\pi
/2 $. As a consequence, the velocity field $\mathbf{v}(\mathbf{r})$
is perfectly symmetric with respect to the interface. In the general case $%
\theta _{0}\neq \pi /2$, the particle center is above or below the
interface. Then the Marongoni flow $\mathbf{v}^{I}$\textbf{\ }is still
symmetric, whereas the contribution arising from the boundary condition at
the particle surface,\textbf{\ }$\mathbf{v}^{P}$, \ is no longer the
same above and below the interface. This would double the number of
independent coefficients in the boundary conditions at the particle surface
and modify the numerical factor in (\ref{29}).\ We do not expect a change of
the qualitative results; in particular the far-field $\mathbf{v}^{I}$ is
not affected by the vertical particle position.

\subsection{Single-particle velocity}

According to (\ref{29}) a non-uniformly heated colloidal sphere moves in the
direction opposite to its self-generated temperature gradient. With $U$ from
Eq. (\ref{15}), the self-propulsion velocity is given by the tension
derivative $\gamma _{T}$, the particle's heat absorption rate $Q$ and dipole
moment $b$, \ and the\ fluid viscosity $\bar{\eta}$ and heat conductivity $%
\bar{\kappa}$. In terms of the excess temperature at the particle surface, $%
\Delta T=Q/2\pi \bar{\kappa}a$, the velocity scale reads as 
\begin{equation}
U=-\frac{\gamma _{T}\Delta T}{2\bar{\eta}}.  \label{36}
\end{equation}%
With the parameters of a water-air interface at room temperature, $\bar{\eta}%
=10^{-3}$ Pa.s and $\gamma _{T}=-2\times 10^{-3}$ NK$^{-1}$m$^{-1}$, we find
that hot particles with an excess temperature $\Delta T=1$ K induce a
Marangoni flow with velocity scale $U$ of the order of $1$ m/s.

The self-propulsion velocity $u$ then depends on the reduced temperature
dipole moment $b/a$. For a particle that is heated through laser-absorbing
molecules dispersed in the particle's bulk material, the asymmetry $b/a$ is
significantly smaller than unity. On the other hand, if the heating occurs
through a metal patch fixed on its surface, one has $b/a\approx 1$ and $%
u\approx -\frac{1}{10}U$. This estimate agrees qualitatively with the
velocity of several cm/s observed by \cite{Oka09} for a macroscopic floater
with a hot spot at one side. Note that, at constant excess temperature, $U$
is independent of the particle size.

\subsection{Non-linear effects}

The stationary temperature profile (\ref{temp}) accounts for diffusive heat
flow from an immobile source but neglects convective transport and the
particle's motion. This approximation is valid for small Marangoni number $%
aU/\alpha $, where $\alpha $ is the thermal diffusivity. For micron-size
particles and $\alpha =1.4\times 10^{-7}$ m$^{2}$/s for water, one find that
the linear approximation ceases to be valid at $U\approx 10$ cm/s. In the
case of a chemical Marangoni effect, the relevant parameter is provided by
the P\'{e}clet number $aU/D$ of the molecular solute with diffusivity $D$.
Since molecules diffuse more slowly than heat ($D$ is about hundred times
smaller than $\alpha $), the linear approximation is restricted to
velocities $U\sim \ $mm/s. Finally, convective acceleration is negligible
for small Reynolds number $aU/\nu $, with the kinematic viscosity $\nu $;
this condition holds true for $U<1$ m/s.

We briefly discuss two non-linear effects, heat advection by the fluid
velocity and distortion of the temperature field due to the particle motion.
The temperature profile of an immobile heat source is determined by $%
\partial _{t}T=\mathbf{\nabla \cdot }(T\mathbf{v}-\alpha \mathbf{\nabla }%
T)$. For small Marangoni number the advective term is negligible, and the
excess temperature varies with the inverse distance. For $aU/\alpha >1$
advection enhances the heat flow and modifies the temperature profile in the
vicinity of the particle ($r<\alpha /U$).

An additional effect occurs at finite particle velocity $u$, where heat
diffusion is efficient at short distances only, thus resulting in strong
retardation effects in the far-field. For a hot particle moving in negative $%
x$-direction, the $1/r$ term of (\ref{temp}) is distorted according to 
\begin{equation}
T(\mathbf{\emph{r}})=T_{0}+\frac{Q}{2\pi \bar{\kappa}r}e^{-(r-x)/\ell },
\end{equation}%
with $\ell =-\alpha /u$. At large distances ($r>\ell $), the diffusive heat
transport does not catch up the particle's motion, resulting in an
asymmetric temperature profile that decays more rapidly in front of the
moving particle ($x<0$) and thus significantly modifies the Marangoni stress
(\ref{2}).

\subsection{Collective effects}

We conclude with a brief discussion of collective effects arising from the
interplay of self-propulsion and hydrodynamic interactions; the latter are
well approximated by the far-field contribution $\mathbf{v}^{I}$. In
addition to the self-generated velocity $\mathbf{u}_{i}=-u\mathbf{e}%
_{i}$, each particle is advected in the Marangoni flow of its neighbors $%
\mathbf{w}_{i}$, resulting in the drift-diffusion equation for the
probability density $\rho (\mathbf{r},\mathbf{e})$, 
\begin{equation}
\partial _{t}\rho +\mathbf{\hat{\nabla}\cdot }(\mathbf{u}+%
\mathbf{w)}\rho =D\mathbf{\hat{\nabla}}^{2}\rho +D_{r}\mathcal{R}%
^{2}\rho ,  \label{30}
\end{equation}%
with the 2D gradient $\mathbf{\hat{\nabla}}$, the rotation operator
about the vertical axis $\mathcal{R}=\mathbf{e}\times \partial _{%
\mathbf{e}}$, and the diffusion coefficients $D$ and $D_{r}$. The
self-generated Marangoni flow of spheres does not exert a viscous torque, in
contrast to sufficiently asymmetric particles \cite{Nak97} and in contrast
to the effective slip velocity in bulk phases \cite{Bic14}; the torque
exerted by the vorticity of the advection flow is small.

Evaluating (\ref{5}) at the interface, one readily obtains the advection
velocity as a gradient field $\mathbf{w}_{i}=-\mathbf{\hat{\nabla}}%
_{i}\Phi _{i}$, where the effective single-particle potential 
\begin{equation}
\Phi _{i}=aU\sum_{j}\left( -\ln r_{ij}+\frac{\mathbf{b}_{j}\cdot 
\mathbf{r}_{ij}}{r_{ij}^{2}}\right) ,  \label{32}
\end{equation}%
depends on the distance vector $\mathbf{r}_{ij}=\mathbf{r}_{j}-%
\mathbf{r}_{i}$ and the temperature dipole $\mathbf{b}_{j}=b%
\mathbf{e}_{j}$ of the neighbor $j$.

For $b\approx 0$ both the self-propulsion velocity $\mathbf{u}$ and the
dipole term in $\Phi _{i}$ are small. For chemically active particles, Eq. (%
\ref{30}) leads to the Keller-Segel model and its rich dynamical behavior 
\cite{Mas14a}. For an external source field, such as the excess temperature
due to laser heating of absorbing particles, an outward Marangoni flow ($U>0$%
) results in a radially symmetric repulsive pair interaction, and the steady
state $\rho _{\text{st}}\propto e^{-\Phi /D}$ is given by the effective
overall potential $\Phi =-aU\sum_{<i,j>}\ln r_{ij}$. This potential favors
an ordered phase of hexatic symmetry, which was indeed observed for camphor
boats floating on water \cite{Soh08}.

A more complex behavior is expected for a non-zero temperature dipole $b$.
During its rotational diffusion time the particle travels over a distance $%
L=u/D_{r}$. If this length is smaller than the nearest-neighbor distance $d$
of the hexatic lattice, self-propulsion merely enhances the effective
diffusion coefficient \cite{How07}. With a rotational diffusion time of $%
1/D_{r}\approx 5$ s, micron-size particles may attain a velocity $u$ of
several mm/s, resulting in a length $L$ of several centimeters. Then the
self-generated motion destroys the hexatic lattice and may lead to
disordered phases where the particle orientations $\mathbf{e}_{i}$ play
the role of quenched random fields.

Acknowledgement. The author thanks Mireille Bousquet-M\'{e}lou, Thomas
Bickel, and Guillermo Zecua for stimulating discussions and helpful remarks.
This work was supported by Agence Nationale de la Recherche through contract
ANR-13-IS04-0003, and by CNRS/IDEX Bordeaux through PEPS \textquotedblleft
Propulsion de micro-nageurs par effet Marangoni\textquotedblright .

\end{document}